# Web Pages Clustering: A New Approach

Jeevan H E [#1], Prashanth P P [#2], Punith Kumar S N [#3], Vinay Hegde [#4]
[#] *Dept. of Computer Science and Engineering, RV College of Engineering, Bangalore, Karnataka, India*

***Abstract*—The rapid growth of web has resulted in vast volume of information. Information availability at a rapid speed to the user is vital. English language (or any for that matter) has lot of ambiguity in the usage of words. So there is no guarantee that a keyword based search engine will provide the required results. This paper introduces the use of dictionary (standardised) to obtain the context with which a keyword is used and in turn cluster the results based on this context. These ideas can be merged with a metasearch engine to enhance the search efficiency.**

***Keywords: Clustering, concept mining, information retrieval, metasearch engine***

## I. INTRODUCTION

As information availability increases with the growth of the web, the number of users who want to retrieve that information also increases. This has led to the rise of search engines. A search engine typically is based on a keyword as a query, uses this to search its indexed database which has data about different web sites and their content and presents the results to the user. But users still find it fairly difficult to find the exact information required by them, even though it may be present in the web. There are various reasons for this.

One reason for this is that many users search the Internet with keywords that are ambiguous to certain degree.

For example : If one searches for "keyboard" in a search engine expecting sites containing information about the musical instrument, he gets a list that is a mix of links to pages containing information about typing keyboard and musical instrument.

Today we have many sophisticated search engines like Google, Yahoo, Bing etc. But still we are not guaranteed of accurate search results. Apart from the above mentioned reason, it may also be due to the fact that a single search engine may not be able to index the entire web which has grown to such a large extent. Every day thousands of new web sites are created and millions of existing pages get updated. To keep track of every such detail is impossible.

In order to solve this problem, many meta-search engines emerge such as, Excite, WebCrawler and so on, which make further processing of search results gathered from many existing search engines as explained in [1]. For example Excite issues queries to three other search engines, including Google, Yahoo, and Bing. The results from these search engines are combined to find the most relevant pages. The advantage is obvious. People can fast identify the information they need.

In this paper we propose a simple and effective method to cluster web pages and extract concepts from a keyword. We also introduce an improved ranking algorithm for metasearch engines.

## II. WEB PAGES CLUSTERING AND CONCEPT MINING

### A. Web Pages Clustering

Clustering can be considered the most important unsupervised learning problem; so, as every other problem of this kind, it deals with finding a structure in a collection of unlabeled data.

A loose definition of clustering could be "the process of organizing objects into groups whose members are similar in some way".

A cluster is therefore a collection of objects which are "similar" between them and are "dissimilar" to the objects belonging to other clusters as defined in [2].

Web pages clustering, in particular, mean removing irrelevant links from the obtained results. The result from multiple search engines is processed to obtain the final search result page. The result which appears in results of more search engines will be listed above the others.

### B. Concept Mining

Concept mining is an activity that results in the extraction of concepts from artefacts. Solutions to the task typically involve aspects of artificial intelligence and statistics, such as data mining and text mining. Because artefacts are typically a loosely structured sequence of words and other symbols (rather than concepts), the problem is nontrivial, but it can provide powerful insights into the meaning, provenance and similarity of documents.

The idea is to use the dictionary available in the Internet to determine the different contexts in which the keyword can appear, that is, the same keyword explaining different concepts.

## III. USE OF DICTIONARY

Concept mining as mentioned earlier involves Artificial Intelligence. Extracting concepts from short text snippets retrieved from the search results may not be accurate enough. To achieve good amount of accuracy, we may require the entire text to be available. Hence it can be computationally intensive and consume high bandwidth to function at an acceptable speed [3]. For the internet environment, a better solution can be to use a dictionary. A dictionary can be used for the queries that the user gives. Each ambiguous word will lead to multiple meanings obtained from the dictionary. Based on these multiple meanings clusters can be done for each type of result.

This clustering can be done in two ways. One is to process the search results. Compare the context of the results with the meanings retrieved from the dictionary. This is again not straightforward and requires considerable data mining techniques [4]. Hence we propose a simple alternative but an efficient technique. The technique is to submit the meanings retrieved itself as queries to the search engine. This eliminates the need for any data mining algorithm. Each result retrieved already belongs to a particular cluster (the meaning used for searching). So this eliminates the need for a clustering algorithm.

Now consider a query such as “Bank”. The dictionary can provide meanings such as financial institution, sides of a water body and rely upon. The search engine can resolve the ambiguity by forming three clusters of results, one for each meaning. The meaning itself is sent to the search engine as a query. Further, the results can be improved by concatenating the user query and the meaning and making it a single new query. In this case it can be “Bank financial institution”.

```
//Module to retrieve meanings from a dictionary
//Input- user query – string
//Output- list of meanings
Dictionary (String query)
do
        meanings = getFromDictionary (query);
        for each meaning from the dictionary
        do
                AddToList (list, meaning)
        end
        if (list is NULL) // no meaning found
                // query may be a noun
                AddToList (list, query)
return list
end
```

When it comes to implementation of the same, the dictionary can be maintained either online or offline. An online dictionary such as that of the WorldNet is a better choice, since it is updated regularly and is widely accepted standard dictionary. On the other hand an offline, local dictionary is also possible, provided it is sophisticated enough to provide the results with minimum delay and can be updated regularly.

One problem with this is the use of multiword queries. In this case, it may still be possible to get the meaning of each word of the query from the dictionary, but constructing a new query from that will be a problem. Different solutions can be provided for the same. The algorithm may be designed to select only one word for querying, based on the number of meanings retrieved for each word in the multiword query. The word with maximum number of different meanings can be used. Another solution is to perform a quick concept mining from the multi word query and obtain a single word query. For Example, a query such as “Where is Bangalore”, can be reduced to just “Bangalore”

Another problem with the use of dictionary is for the queries that involve proper nouns. The dictionary is not expected to provide results for these. Even proper nouns can be ambiguous to some extent. For example consider “Sachin”. This could refer to cricket player Sachin Tendulkar or any other individual with the same name ( music director Sachin Dev Burman), resolving such ambiguities is non-trivial and may require more input from the user itself. One approach to remove such ambiguities is to use the history of searches by the same user [5]. This can inherently point to a certain context. In this case if the user had earlier searched things about sports, then the probability is more that the query “Sachin” meant “Sachin Tendulkar”. This requires data mining and statistical analysis of previous data available.

## IV. METASEARCH ENGINE

A metasearch engine is a search tool that sends user requests to several other search engines and/or databases and aggregates the results into a single list and provides it to the user in way similar to any other search engine. The concept of metasearch engine arises from the fact that the web is too large for one search engine to index it completely and more comprehensive results can be obtained by combining the results of various search engines [6]. The obvious advantage of this technique is that the search space is more i.e. more web pages are covered. Since a metasearch engine has to deal with different search engines, it requires a parsing stage to convert the results from all the search engines into a uniform manner. The implementation can typically involve XML and HTML parsing.

The usage of a metasearch engine must be done in an intelligent manner to extract the maximum benefit out of it. The ranking of results is very crucial to provide the user with the required information in minimum time. A straightforward algorithm that can be adopted to provide a well refined search result is given below. The underlying assumption is that a few results will be same, from all the search engines. Here we consider the count of each result link from all the search engines used. Then rank it, based on the decreasing order of the count.

```
    //Module to search and unify the results
    //Performs ranking based on the count
    //Input: user query – string
    //Output: list of browsable search results

MetaSearchEngine (query)
do
    Submit the search query to the search engines
        for each search engine
        do
            for each result_link from the given
                search engine
            do
            if (Final_Results has result_link)
            // increment count
            SetCount (result_link, getCount
            (result_link) +1)
            else
            //add it to result list and set count to 1
            AddToFinalResults(result_link)
            SetCount(result_link, 1)
            end
        end
    Sort the Final_Results in the decreasing order of the
    count of the result
    Display the search results in this order

end
```

The use of this approach provides a far more efficient ranking than simply performing a union of all the results. Moreover it's a simple approach and easily implementable. This ranking can also be done on client side (using client side scripting). Hence it provides a flexible approach for implementation. Experimental implementation of the same technique has been done, with a good amount of success.

## V. CONCLUSION

The paper proposed a new basis for web pages clustering and concept extraction from a keyword based on results of multiple search engines on the Internet. It will help user to get relevant information needed upon querying. We also did an experimental implementation of the same ideas, which performed to meet our expectations of speed and efficiency.

It can be said that providing context sensitive results increases the efficiency of the user, so that he can easily find the document he is searching for in the web.

Current keyword based search engines rank the web pages based on frequency of the keywords, inbound link count etc. Hence these results require user to go through all the returned links for finding the right one. With the use of a metasearch engine the relevance of results is also high, since it uses multiple search engines like Google, Yahoo and Bing. The links that appear in most of search engines' results are given higher priority.

Further enhancements include support for queries from languages other than English, enabling caching mechanism for recently queried keywords and moving forward to implement the above idea for image searching as well as video searching.

## ACKNOWLEDGMENT

We would like to thank. Dr. T. M. Rangaswamy, Professor, IEM department, R.V College of Engineering for providing support and guidance for the study and research regarding the subject.

# A Performance Study of Data Mining Techniques: Multiple Linear Regression vs. Factor Analysis

**Abhishek Taneja, R.K.Chauhan**
Assistant Professor, Dept. of Computer Sc. & Applications, DIMT, Kurukshetra
Professor, Dept. of Computer Sc. & Applications,Kurukshetra University, Kurukshetra

***Abstract:**The growing volume of data usually creates an interesting challenge for the need of data analysis tools that discover regularities in these data. Data mining has emerged as disciplines that contribute tools for data analysis, discovery of hidden knowledge, and autonomous decision making in many application domains. The purpose of this study is to compare the performance of two data mining techniques viz., factor analysis and multiple linear regression for different sample sizes on three unique sets of data. The performance of the two data mining techniques is compared on following parameters like mean square error (MSE), R-square, R-Square adjusted, condition number, root mean square error(RMSE), number of variables included in the prediction model, modified coefficient of efficiency, F-value, and test of normality. These parameters have been computed using various data mining tools like SPSS, XLstat, Stata, and MS-Excel. It is seen that for all the given dataset, factor analysis outperform multiple linear regression. But the absolute value of prediction accuracy varied between the three datasets indicating that the data distribution and data characteristics play a major role in choosing the correct prediction technique.*

**Keywords: Data mining, Multiple Linear Regression, Factor Analysis, Principal Component Regression, Maximum Liklihood Regression, Generalized Least Square Regression**

## 1. Data Introduction

A basic assumption concerned with general linear regression model is that there is no correlation (or no multi-collinearity) between the explanatory variables. When this assumption is not satisfied, the least squares estimators have large variances and become unstable and may have a wrong sign. Therefore, we resort to biased regression methods, which stabilize the parameter estimates [17]. The data sets we have chosen for this study have a combination of the following characteristics: few predictor variables, many predictor variables, highly collinear variables, very redundant variables and presence of outliers.

The three data sets used in this paper viz., marketing, bank and parkinsons telemonitoring data set are taken from [8],[9], and [10] respectively.

From the foregoing, it can be observed that each of these three sets has unique properties. The marketing dataset consists of 14 demographic attributes. The dataset is a good mixture of categorical and continuous variables with a lot of missing data. This is characteristic for data mining applications.

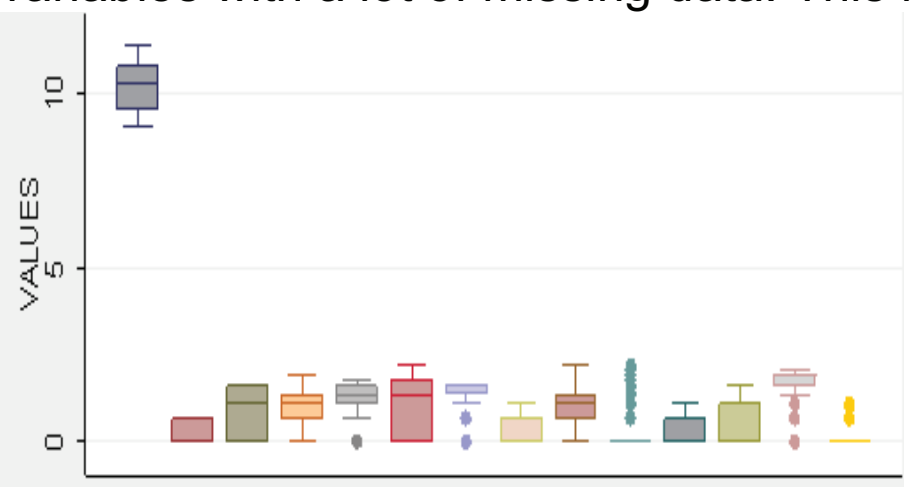

Fig 1 Box Plot of Marketing Dataset

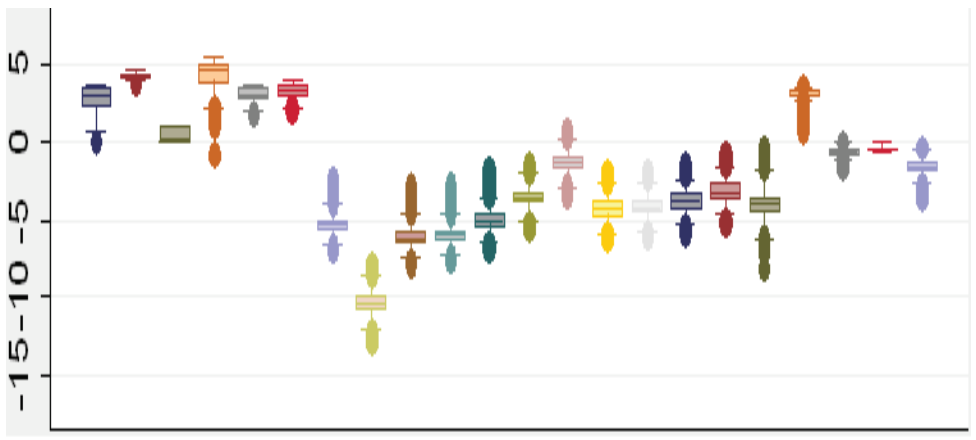

Fig 2: Box Plot of Parkinson Dataset

The bank dataset is synthetically generated from a simulation of how bank-customers choose their banks. Tasks are based on predicting the fraction of bank customers who leave the bank because of full queues.

Each bank has several queues, that open and close according to demand. The tellers have various affectivities, and customers may change queue, if their patience expires.

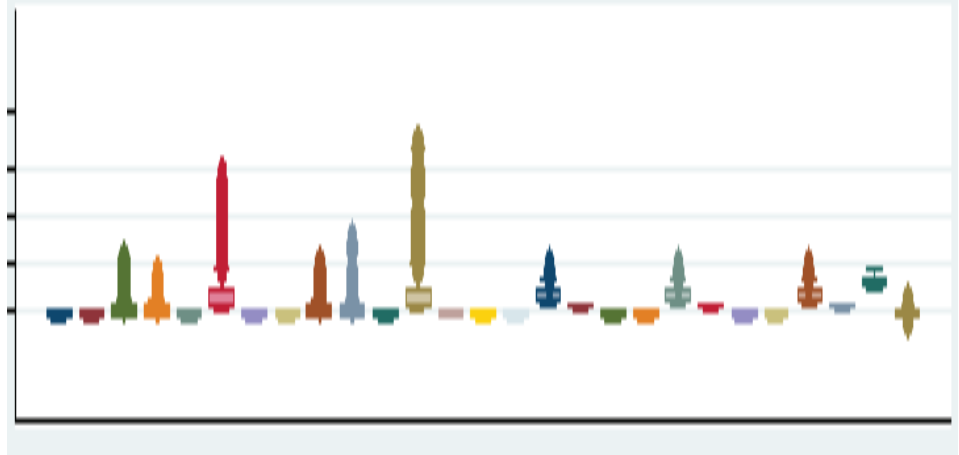

Fig 3: Box Plot of Bank Dataset

In the rej prototasks, the object is to predict the rate of rejections, i.e., the fraction of customers that are turned away from the bank because all the open tellers have full queues. This dataset consists of 32 continuous attributes and having 4500 records.
The parkinsons telemonitoring data set is composed of a range of biomedical voice measurements from 42 people with early-stage Parkinson's disease recruited to a six-month trial of a telemonitoring device for remote symptom progression monitoring. The recordings were automatically captured in the patient's homes. Columns in the table contain subject number, subject age, subject gender, time interval from baseline recruitment date, motor UPDRS, total UPDRS, and 16 biomedical voice measures. Each row corresponds to one of 5,875 voice recording from these individuals. The main aim of the data is to predict the total UPDRS scores ('total_UPDRS') from the 16 voice measures. This is a multivariate dataset with 26 attributes and 5875 instances. All the attributes are either integer or real with lots of missing and outlier values.
The box plot of the three datasets (fig 1 to fig.3) shown above display measure of dispersion between these variables, compares the mean of different variables, and also shows the outliers in three datasets. In this regard, it becomes necessary to scale these three datasets to reduce the measure of dispersion and bring all the variables of all datasets to the same unit of measure.

## 2. Prediction Techniques

There are many prediction techniques (association rule analysis, neural networks, regression analysis, decision tree, etc.) but in this study only two linear regression techniques have been compared.

### 2.1 Multiple Linear Regression

Multiple linear regression model maps a group of predictors x to a response variable y [4]. The multiple linear regression is defined by the following relationship, for *i* = 1*,* 2*, n*:

$$y_i = a + b_1x_{i1} + b_2x_{i2} + \cdot \quad \cdot \quad \cdot + b_kx_{ik} + e_i$$

or, equivalently, in more compact matrix terms:

$$Y = Xb + E$$

where, for all the *n* considered observations, ***Y*** is a column vector with *n* rows containing the values of the response variable; **X** is a matrix with *n* rows and *k* + 1 columns containing for each column the values of the explanatory variables for the *n* observations, plus a column (to refer to the intercept) containing *n* values equal to 1; ***b*** is a vector with *k* + 1 rows containing all the model parameters to be estimated on the basis of the data: the intercept and the *k* slope coefficients relative to each explanatory variable. Finally ***E*** is a column vector of length *n* containing the error terms. In the bivariate case the regression model was represented by a line, now it corresponds to a (*k* + 1)-dimensional plane, called the regression plane. This plane is defined by the equation

$$\hat{y}_i = a + b_1x_{i1} + b_2x_{i2} + \cdot \quad \cdot \quad \cdot + b_kx_{ik} + \mu_i$$

Where $\hat{y}_i$ is dependent variable. $X_i^{'s}$ are independent variables, and $\mu_i$ is stochastic error term. We have compared three basic methods under this multiple linear regression technique. They are full method (which uses the least square approach), forward method, and stepwise approach (which used discriminant approach or all possible subsets) [5].

### 2.2 Factor Analysis

Factor analysis attempts to represent a set of observed variables $X_1$, $X_2$ .... $X_n$ in terms of a number of 'common' factors plus a factor which is unique to each variable. The common factors (sometimes called latent variables) are hypothetical variables which explain why a number of variables are correlated with each other- it is because they have one or more factors *in common* [7].

Factor analysis is basically a one-sample procedure [6]. We assume a random sample $\mathbf{y}_1$, $\mathbf{y}_2$, $\mathbf{y}_n$ from a homogeneous population with mean vector $\mu$ and covariance matrix $\Sigma$. The factor analysis model expresses each variable as a linear combination of underlying *common factors* $f_1, f_2, \ldots, f_m$, with an accompanying error term to account for that part of the variable that is unique (not in common with the variables). For $y_1$, $y_2$, $y_p$ in any observation vector $\mathbf{y}$, the model is as follows:

$$y_1 - \mu_1 = \lambda_{11} f_1 + \lambda_{12} f_2 + \cdots + \lambda_{1m} f_m + \varepsilon_1$$

$$y_2 - \mu_2 = \lambda_{21} f_1 + \lambda_{22} f_2 + \cdots + \lambda_{2m} f_m + \varepsilon_2$$

$$\ldots$$

$$y_p - \mu_p = \lambda_{p1} f_1 + \lambda_{p2} f_2 + \cdots + \lambda_{pm} f_m + \varepsilon_p.$$

Ideally, $m$ should be substantially smaller than $p$; otherwise we have not achieved a parsimonious description of the variables as functions of a few underlying factors. We might regard the *f's* in equations above as random variables that engender the $y$'s. The coefficients $\lambda_{ij}$ are called *loadings* and serve as weights, showing how each $y_i$ individually depends on the $f$'s. With appropriate assumptions, $\lambda_{ij}$ indicates the importance of the $j$th factor $f_j$ to the $i$th variable $y_i$ and can be used in interpretation of $f_j$. We describe or interpret $f_2$, for example, by examining its coefficients, $\lambda_{12}$, $\lambda_{22}$, $\lambda_{p2}$. The larger loadings relate $f_2$ to the corresponding $y$'s. From these $y$'s, we infer a meaning or description of $f_2$. After estimating the $\lambda_{ij}$'s, it is hoped they will partition the variables into groups corresponding to factors. There is superficial resemblance to the multiple linear regression, but there are fundamental differences. For example, firstly f's in above equations are unobserved, secondly equations above represents one observational vector, whereas multiple linear regression depicts all n observations.

There are a number of different varieties of factor analysis: the comparison here is limited to principal component analysis, generalized least square and maximum likelihood estimation.

## 3. Related Work

There are many data mining techniques (decision tree, neural networks, regression, clustering etc.) but in this paper we have compared two linear techniques viz., multiple linear regression, and factor analysis. In this domain there have been many researchers and authors who compared various data mining techniques from varied aspects.

In year 2004 Munoz et. al did a comparison of three data mining methods: linear statistical methods, neural network method, and non-linear multivariate methods [11]. In 2008, Saikat and Jun Yan compared PCA and PLS on simulated data [12]. Munoz et.al compared logistic regression, principal component regression, and classification and regression tree with multivariate adaptive regression spines [16]. In 1999, Manel et.al compared discriminate analysis, neural networks, and logistic regression for predicting species distribution [13]. In year 2005, Orsalya et. al compared ridge regression, pair wise correlation method, forward selection, best subset selection, on quantitative structure retention relationship study based on multiple linear regression on predicting the retention indices for aliphatic alcohols[14]. In year 2002 Huang et. al compared least square regression, ridge and partial least square in the context of the varying calibration data size using only squared prediction errors as the only model comparison criteria [15].

## 4. Preparation and Methodology

Both the techniques under study are linear in nature and the choice of technique is vital for getting significant results. When a nonlinear data are fitted to a linear technique, the results obtained are biased and when linear data are fitted to a non-linear technique, the results have increased variance. As the techniques undertaken for this study are both linear, so to get significant results we need to apply the same on linear data sets. Both the techniques are linear regression techniques, we mean that they are linear in parameters [1] [2]; the $\beta$'s (that is, the parameters are raised to the first power only. It may or may not be linear in explanatory variables, the X's. To make our data sets linear it is preprocessed by taking natural log of all the instances of the data sets or normalized using z-score [3] normalization. After scaling and standardizing the three datasets, it is found that skewness is reduced that is shown by histogram diagram of all three datasets. For proving linearity of these data sets box-plot, histogram and JB Test (Jarque Bera Test) with p-value (exact significance level or probability value of committing type-I error) have been used.
After scaling and standardizing the data sets are divided into two parts, taking 70% observations as the "training set" and the remaining 30% observations as the "test validation set"[3]. For each data set training set is used to build the model and various methods of that technique are employed. For example in Multiple Linear Regression (MLR), three methods are associated in this study: the full model, forward model and stepwise model. The model is validated using test validation data set and the results are presented using ten goodness of fit criteria. Both the techniques are intra and inter compared for their performance on the underlying three unique datasets.

## 5. Interpretation and Findings

Refer to table 1and table 2 given below.

### 5.1 Interpreting Marketing Dataset

In marketing dataset, the value of $R^2$ and Adj.$R^2$, of full model was found with good explanatory power i.e., 0.47, which is higher than both stepwise and forward model.

On the behalf of this explanatory power value we can say that among all methods of multiple linear regression, full model was found best method for data mining purpose, since 47% change in variation in dependent variable was explained by independent variables. But 0.47 value of explanatory power is not significant up-to the mark which requires another regression model than multiple regression model for reporting data set, since 0.53 means 53% of the total variation was found unexplained. So, within multiple regression techniques full model was found best but not up-to the mark. Value of $R^2$ suggest for using another regression model.

| | Methods | MSE | MAE | CN | No. of variables | R Square | Adj. R Square | RMSE | F Value (dF, No. of Observations) | Modified Coefficient of efficiency | Test of normality |
|---|---|---|---|---|---|---|---|---|---|---|---|
| MLR (MARKETING DATASET) | FULL MODEL | 0.333 | 0.33 | 6.87e+6 | 13 | 0.4765 | 0.4751 | .57728 | 336.50 (13, 4805) | -0.009 | 0.6325 |
| | STEPWISE MODEL | 0.603 | 4.94 | 5.10e+5 | 13 | 0.436 | 0.435667 | 0.77 | 1042.32 (11,4805) | 0.047 | 0.6162 |
| | FORWARD MODEL | .584 | 0.897 | 3.53e+3 | 13 | .459 | .458 | 0.76 | 410.48 (13,4805) | 0.077 | 0.6826 |
| MLR (PARKINSON DATASET) | FULL MODEL | 1.256 | 18.55 | 8.284e+8 | 19 | 0.9073 | 0.9068 | .13359 | 2106.68 (19, 4092) | 56.10 | 0.7251 |
| | STEPWISE MODEL | 0.021 | 9.936 | 8.67e+8 | 19 | .910 | .909 | 0.144 | 2288.954 (18,4092) | 0.090 | 0.7343 |
| | FORWARD MODEL | .171 | 10.04 | 0.607e+.6 | 19 | 0.196 | 0.193 | 0.42 | 62.351 (16,4092) | 0.139 | 0.7651 |
| MLR (BANK DATASET) | FULL MODEL | 3.818 | 2.534 | 0.33212 | 32 | 0.0348 | 0.0248 | 1.9542 | 3.51 (32, 3116) | 6.786 | 0.7876 |
| | STEPWISE MODEL | 4.54 | 2.865 | 0.4534 | 32 | 0.0563 | 0.0527 | 2.1307 | 4.45 (32, 3116) | 7.896 | 0.765 |
| | FORWARD MODEL | 4.86 | 2.476 | 0.4653 | 32 | 0.0564 | 0.05383 | 2.204 | 3.851 (32,3116) | 6.765 | 0.5876 |

**Table 1**

The inclusion of some other independent variables (either relevant or irrelevant) in multiple regression model mostly generate non-decreasing explanatory value or $R^2$ value. In this case we can use anther good measure of $R^2$ i.e., Adj. $R^2$, which accounts for the effect of new explanatory variables in the model, since it incorporate degree of freedom of the model, or denominator of the explained and unexplained variation[18]. The expression for the adjusted multiple determination is:

$$\text{Adj. } R^2 = 1-(1-r^2)\ \frac{n-1}{n-k}$$

$$\text{Adj. } R^2 = 1- \left[\frac{\sum e_i^2/(n-k)}{\sum y^2/(n-1)}\right]$$

If n is large Adj. $R^2$ and $R^2$ will not differ much. But with small samples, if the number of regressors X's is large in relation to the sample observations Adj. $R^2$ will be much smaller than $R^2$ and can even assume negative values in which case Adj. $R^2$ should be interpreted as being equal to zero.

For marketing data set, all methods of multiple linear regression Adj. $R^2$ was found similar to $R^2$ value which means sample size is sufficiently large as required for data mining purpose [19].

| | Methods | MSE | MAE | CN | No. of variables | R Square | Adj. R Square | RMSE | F-Value (dF, No. of Observations) | Modified Coefficient of efficiency | Test of normality |
|---|---|---|---|---|---|---|---|---|---|---|---|
| FACTOR ANALYSIS (*MARKETING DATASET*) | PCR | 0.756 | 3.67 | 12 | 13 *(with four components)* | 0.584 | 0.56 | 0.8694 | 323.65 (13,4819) | 5.754 | 0.6654 |
| | MAXIMUM LIKLIHOOD | 0.775 | 3.98 | 9.78e+9 | 13 | 0.589 | 0.576 | 0.8803 | 367.455 (13,4819) | 5.9876 | 0.6792 |
| | GLS | 0.746 | 3.998 | 11 | 13 | 0.587 | 0.573 | 0.8602 | 386.78 (13,4819) | 5.7685 | 0.6776 |
| FACTOR ANALYSIS (*PARKINSON DATASET*) | PCR | 0.456 | 0.67 | 7.87e+7 | 19 *(with six components)* | 0.63 | 0.51 | 0.6749 | 543.5 (19,4112) | 8.56 | 0.87 |
| | MAXIMUM LIKLIHOOD | 0.582 | 0.655 | 7.10e+7 | 19 | 0.64 | 0.54 | 0.763 | 513.65 (19,4112) | 9.38 | 1.73 |
| | GLS | 0.398 | 1.677 | 6.54e+6 | 19 | 0.67 | 0.56 | 0.63 | 665.45(11, 4112) | 11.09 | 1.96 |
| FACTOR ANALYSIS (*BANK DATASET*) | PCR | 0.643 | 0.58 | 8.86e+8 | 33 *(with six components)* | 0.74 | 0.69 | 0.80 | 654.45 (34,3150) | 0.0544 | 0.6758 |
| | MAXIMUM LIKLIHOOD | 0.665 | 0.598 | 8.75e+8 | 33 | 0.728 | 0.684 | 0.815 | 675.65 (34,3150) | 0.0546 | 0.0754 |
| | GLS | 0.678 | 0.612 | 8.74e+8 | 33 | 0.715 | 0.682 | 0.823 | 688.45 (34,3150) | 0.0568 | 0.0543 |

**Table 2**

The $R^2$ in case of marketing dataset for factor analysis was found around 0.58. So, all methods have equal explanatory power under factor analysis. More over, under all methods viz., PCR, Maximum Likelihood, and GLS, explained variation is 58% out of total variation in the dependent variable which signifies that factor analysis extraction is better than multiple linear regression. $R^2$ can also be estimated through the following notations: $R^2 = \frac{ESS}{TSS}$

TSS = Explained Sum Square(ESS)+ Residual Sum Square(RSS)

The Adj. $R^2$ i.e., adjusted for inclusion of new explanatory variable was also found 0.56 less than $R^2$. The 58% variation was captured due to regression, it explains the overall goodness of fit of the regression line to marketing dataset due to use of factor analysis.

So, on the behalf of first order statistical test ($R^2$), we can conclude that factor analysis technique is better than multiple regression technique due to explanatory power.

Mean Square Error (MSE) criteria is a combination of unbiased-ness and the minimum variance property. An estimator is a minimum MSE estimator if it has smallest MSE, defined as the expected value of the squared differences of the estimator around the true population parameter b. MSE( $\hat{b}$ ) =E( $\hat{b}$-b)$^2$ . It can be proved that it is equal to

$$\text{MSE}(\hat{b})\text{'s}$$
$$=\text{Var}(\hat{b})\text{'s}+\text{bias}^2(\hat{b})$$

The MSE criteria for unbiased-ness and minimum variance were found increasing under multiple linear regression models. It signifies that full method MSE is less than all model's MSE, which further means that under full model of multiple linear regression of marketing dataset there is less unbiased-ness and less variance.

The minimum variance also increases the probability of unbiased-ness and gives better explanatory power like $R^2$ in marketing dataset.

The inter comparison of two techniques multiple linear regression and factor analysis generated that in factor analysis models MSE is significantly different which signifies that under factor analysis all b's are unbiased but with large variance. Due to large variance in factor analysis techniques the probability value of unbiased-ness increases that generates a contradictory result about the explanatory power of the factor analysis methods. But factor analysis methods may have questionable values of MSE, due to this reason new measure of MSE that is RMSE (root mean square error) was used in the study.

RMSE was found considerably similar in methods of both the techniques. Due to less variation in RMSE of both MLR and factor analysis of marketing dataset it can be stated that both techniques have equal weights for consideration.

A common measure used to compare the prediction performance of different models is Mean Absolute Error (MAE).

If $Y^p$ be the predicted dependent variable and Y be the actual dependent variable then the MAE can be computed by

$$\text{MAE}=\frac{1}{n}\frac{\sum\left|Y-Y^p\right|}{Y}$$

In marketing dataset MAE was found less under full model, which is less than stepwise and forward model. MAE signifies that full model under MLR techniques give better prediction than other mode

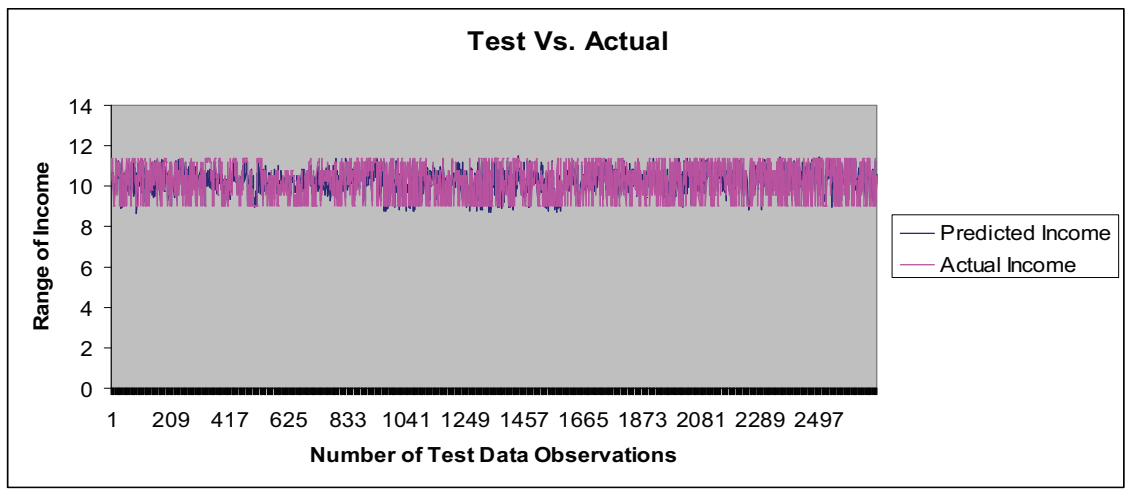

Fig 4: MLR-Full Model (Marketing)

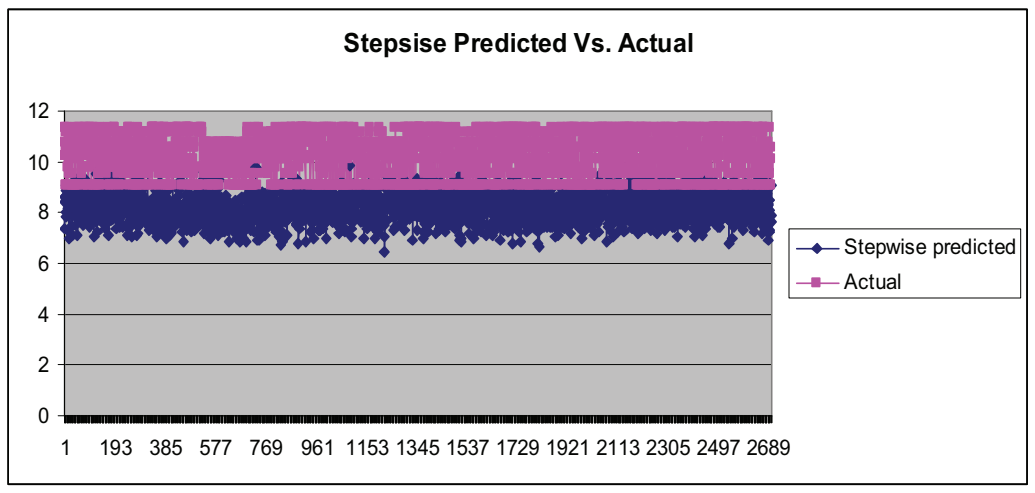

Fig 5: MLR-Stepwise Model (Marketing)

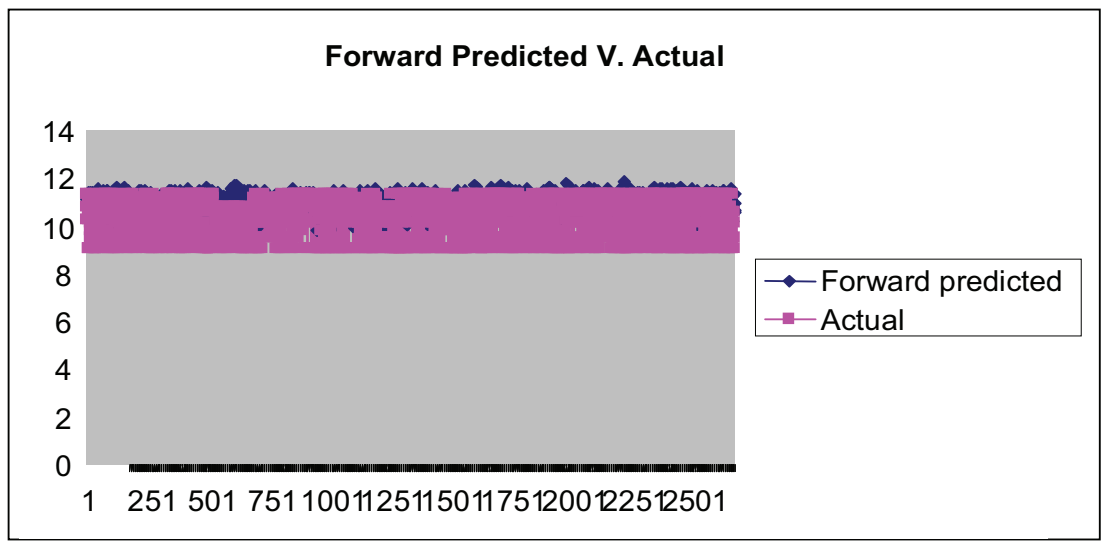

Fig 6: MLR-Forward Model (Marketing)

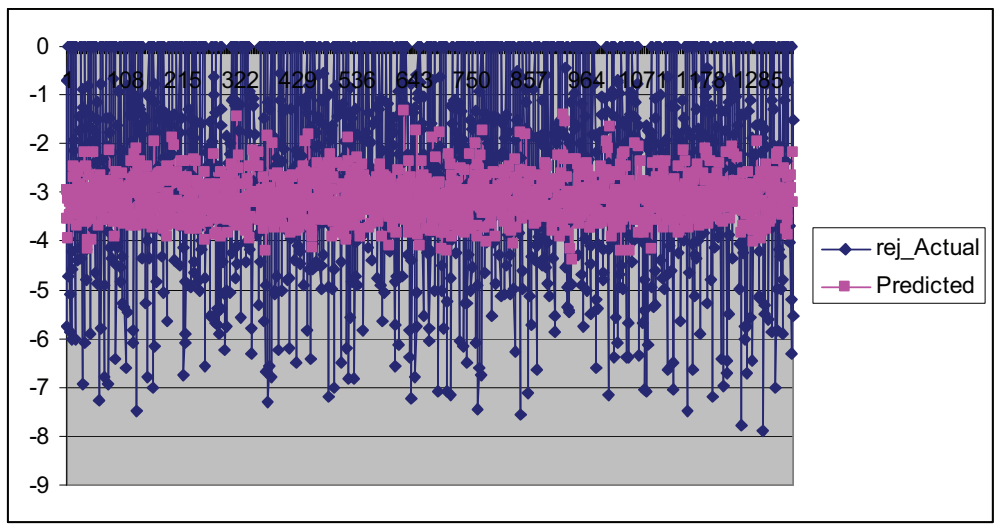

Fig 7: MLR-Full Model (Bank Dataset)

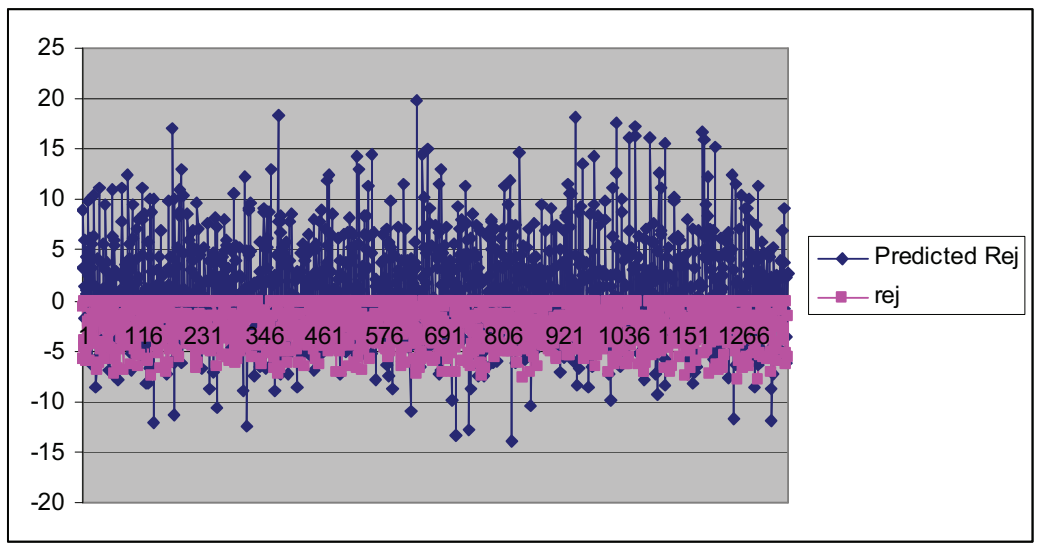

Fig 8: MLR-Forward Model (Bank Dataset)

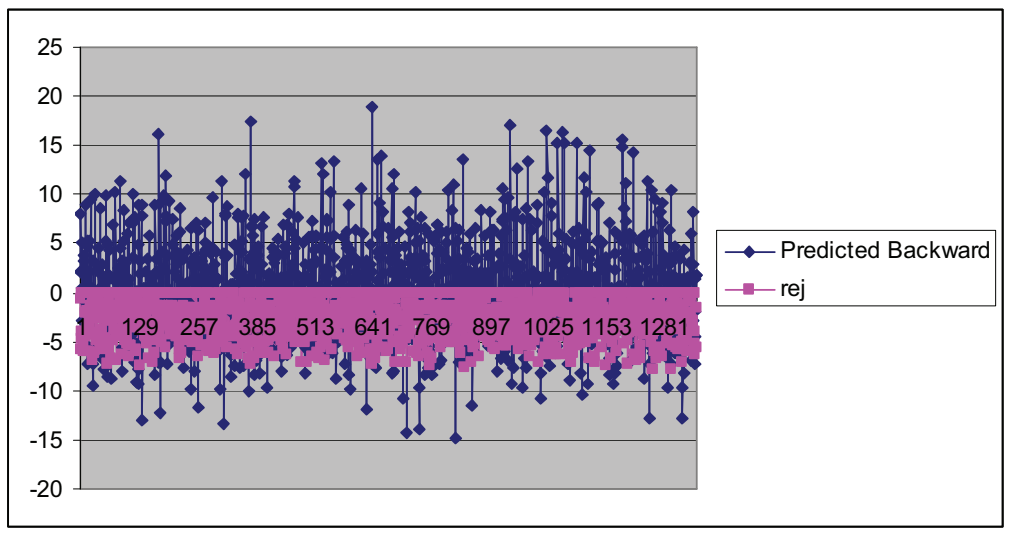

Fig 9: MLR-Stepwise Model (Bank Dataset)

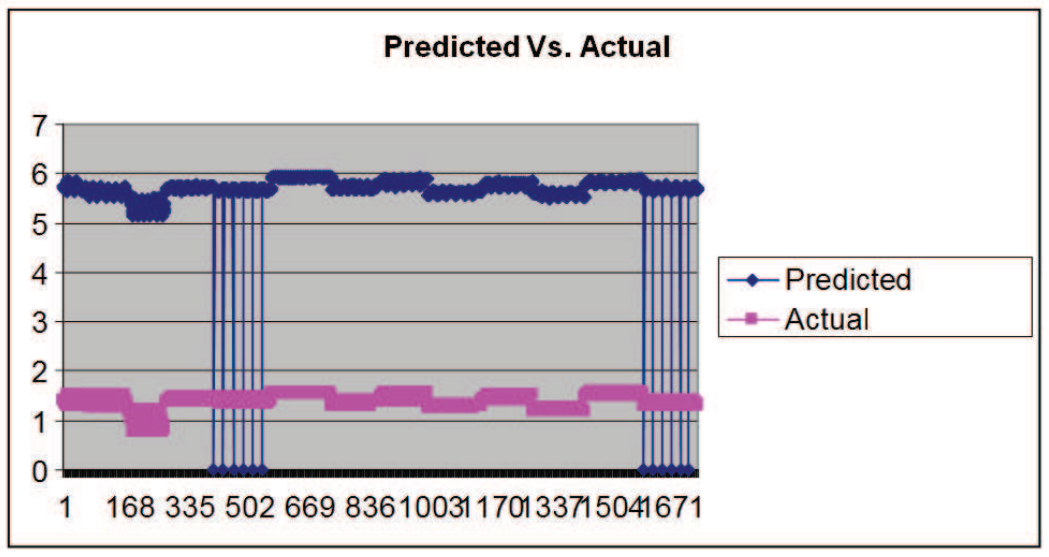

Fig 10: MLR-Full Model (Parkinson Dataset)

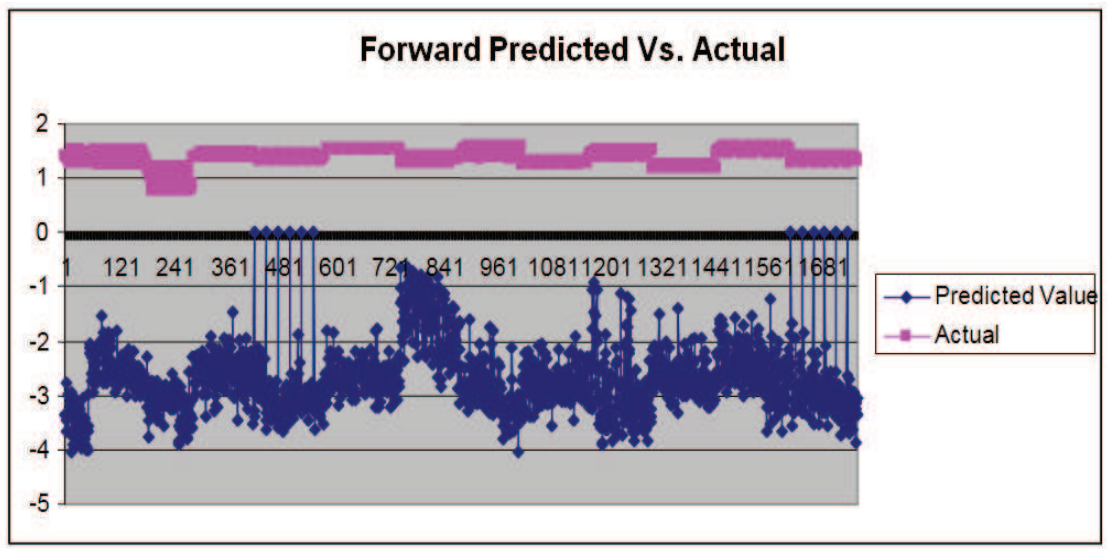

Fig 11: MLR-Forward Model (Parkinson Dataset)

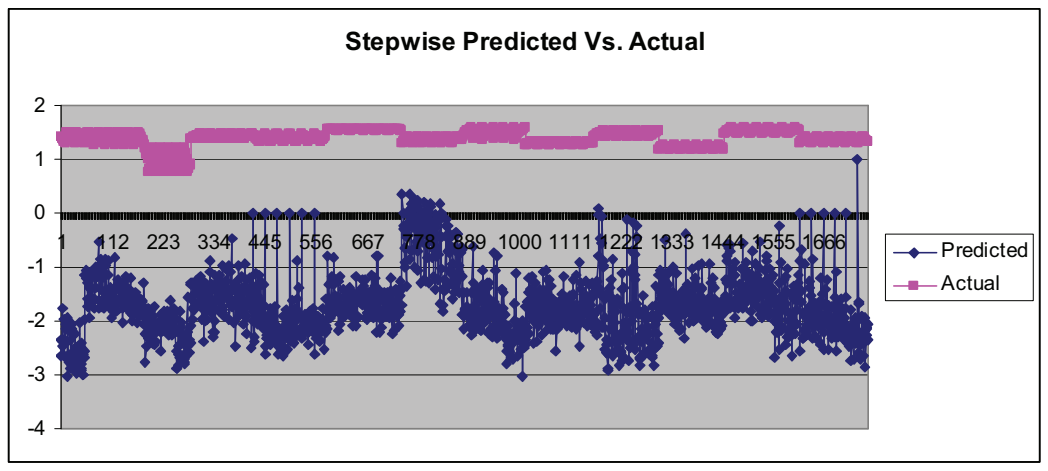

Fig 12: MLR-Stepwise Model (Parkinson Dataset)

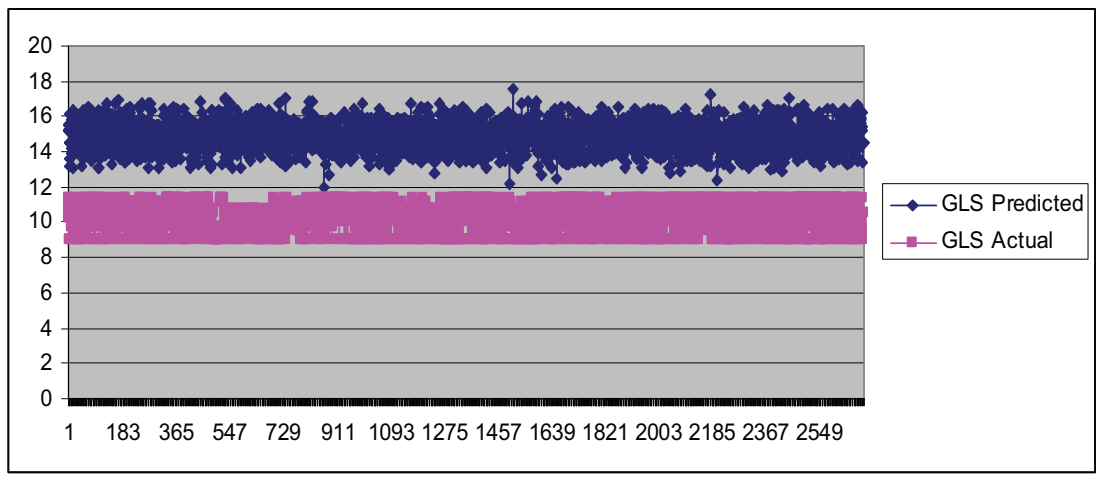

Fig 13: Factor Analysis-GLS Model (Marketing Dataset)

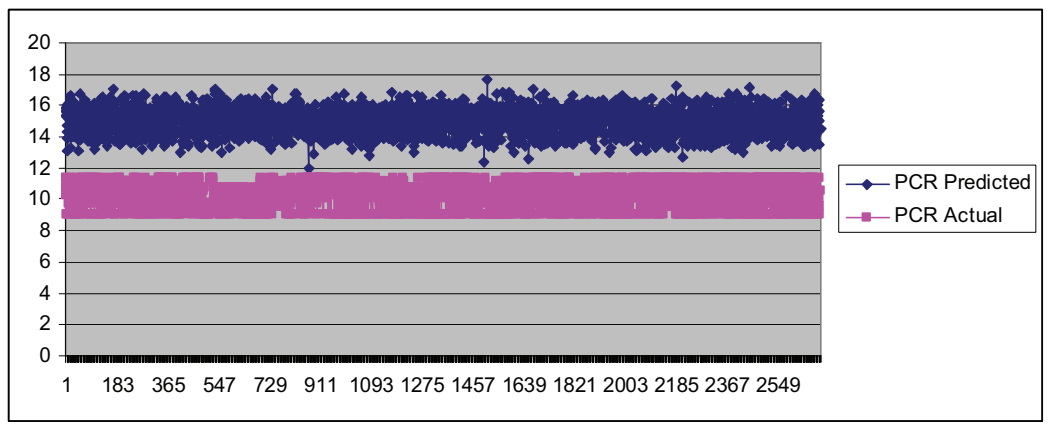

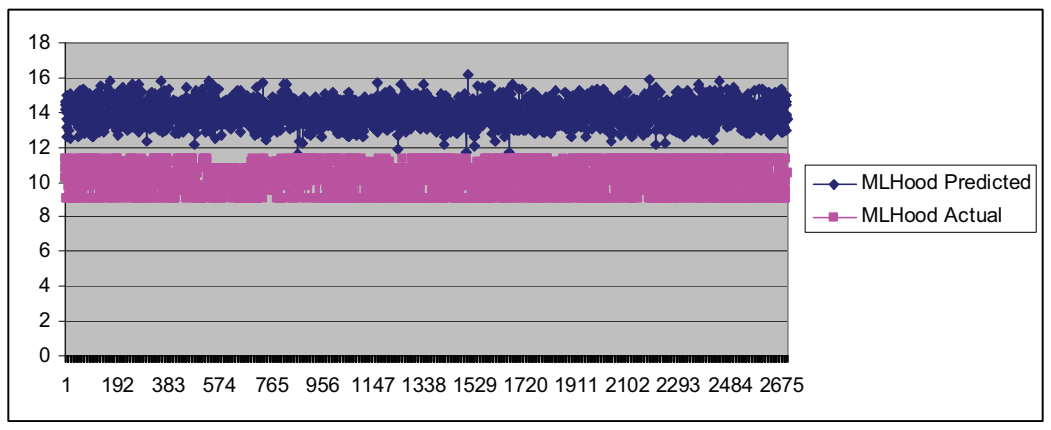

Fig 14: Factor Analysis-PCR Model (Marketing Dataset) Fig 15: Factor Analysis-Maximum Likelihood Model (Marketing Dataset)

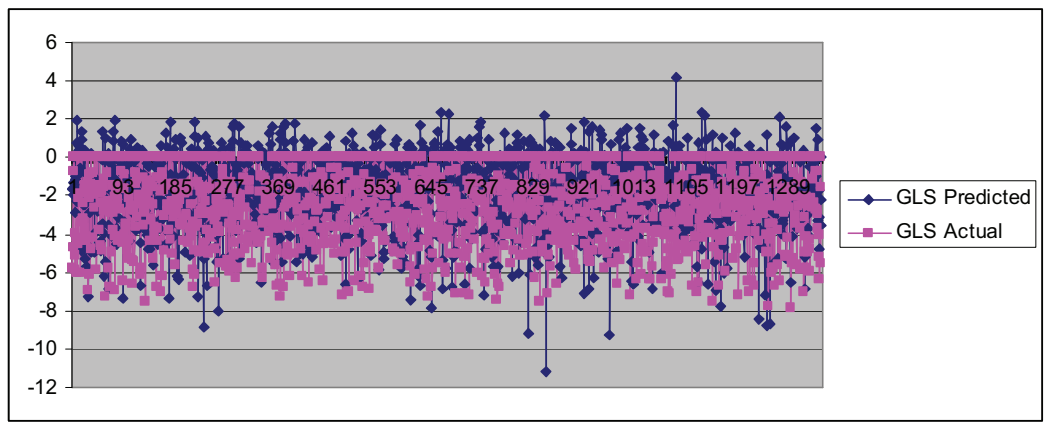

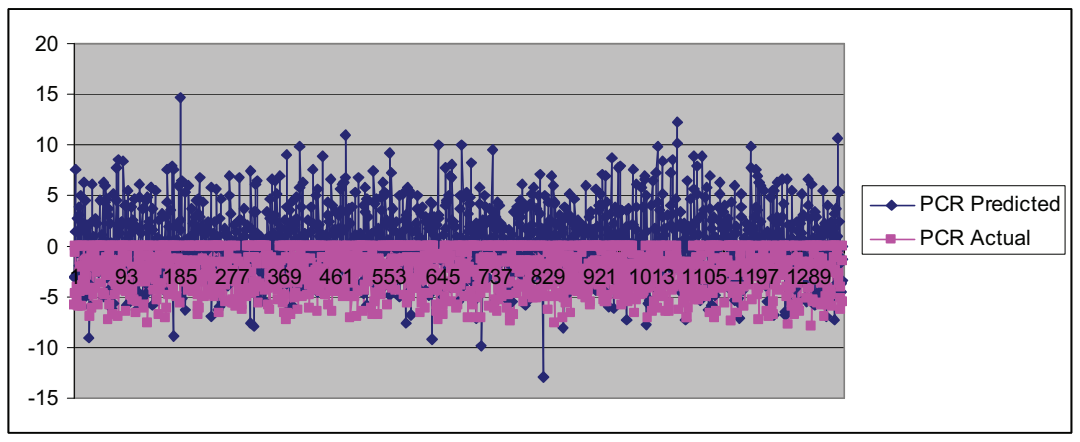

Fig 16: Factor Analysis-GLS Model (Bank Dataset)

Fig 17: Factor Analysis-PCR Model (Bank Dataset)

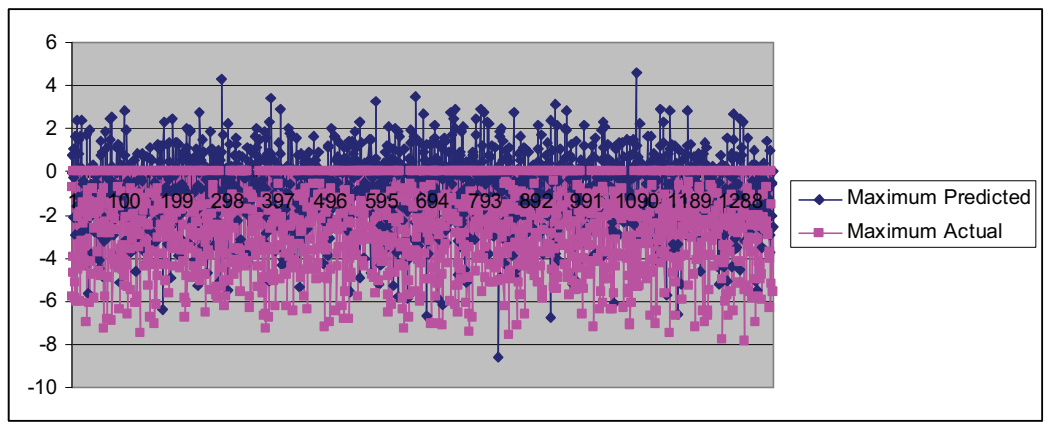

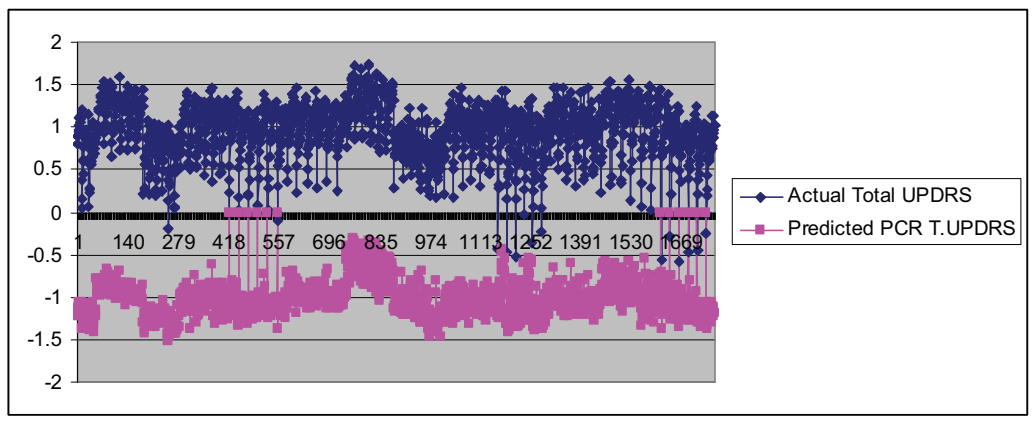

Fig 18: Factor Analysis-Maximum Likelihood Model (Bank Dataset)

Fig 19: Factor Analysis-PCR Model (Parkinson Dataset)

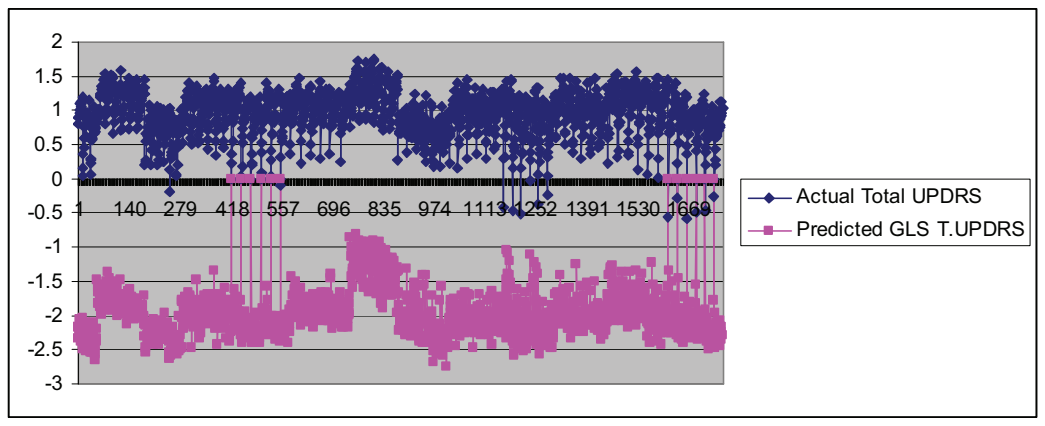

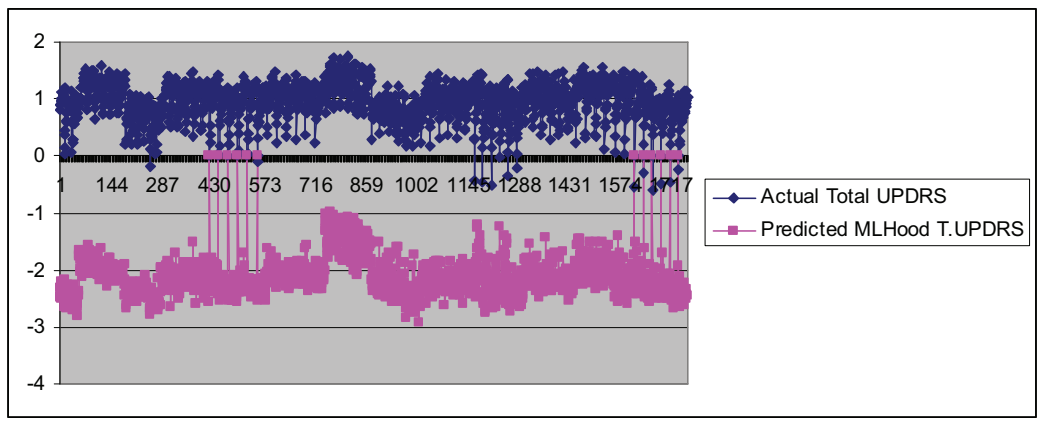

Fig 20: Factor Analysis-GLS Model (Parkinson Dataset)

Fig 21: Factor Analysis-Maximum Likelihood Model (Parkinson Dataset)

Under factor analysis marketing dataset MAE in all models was found considerably similar but higher than multiple regression techniques, therefore we can say factor analysis models for such kind of datasets generate poor prediction performance.

The diagnosis index of multi collinearity was found significantly below 100 under MLR methods in marketing dataset, which means there is no scope for high and severe multi collinearity. In case of same dataset condition number was found lower than factor analysis technique. This means factor analysis is better technique to diagnosis the effect of multi collinearity. But in marketing dataset both factor analysis and MLR techniques were found with less multi collinearity in regressors than severe level of multi collinearity.

The F value in case of marketing dataset was found more than critical value with respect to dF(degree of freedom), in both techniques, which signifies that overall regression model is significantly estimated but stepwise model of MLR technique was found high F corresponding to its dF which means overall significance of the regression model was up-to the mark in case of stepwise method. The prediction plots of two techniques on marketing dataset better represent above discussion visually (see fig. 4-fig. 6 and fig. 13- fig. 15)

### 5.2 Interpreting Bank Dataset

In case full model of bank dataset explanatory power ($R^2$) was found considerably low due to residual, whereas in stepwise and forward model MLR generated satisfactory explanatory power. Due to stepwise and forward model 56% variation in dependent variable was explained with respect to independent variables. Another measure of explanatory power was also found satisfactory in case of stepwise and forward model but not in full model.

On the other hand factor analysis models on bank dataset generated higher value of both $R^2$ and adjusted $R^2$, which signifies that the explanatory power of factor analysis in case of bank dataset is more than MLR technique. Overall one drastic point was found that in all models of factor analysis and MLR, full model of MLR generated very poor $R^2$ value, which means this dataset is not having proper specification according to magnitude change.

The MSE criteria for unbiasness and minimum variance for all parameters is found increasing under both factor analysis and MLR techniques, but all models of factor analysis are found with low unbiasness and variance than all models of MLR. It means both the technique parameters are significant, but MLR techniques parameters are significant with high variance.

The RMSE is also satisfactory and upto the mark in case of factor analysis. Therefore, we can say that factor analysis parameters have low variance and unbiasness.

The prediction power of the regression model is also found good fit in all factor analysis models. In case of bank dataset MLR is having more MAE due to test dataset skewness.

Modified coefficient of efficiency was found low in case of factor analysis model in case of bank dataset, since this dataset does not satisfy the center limit theorem due to constant number of variables; but in MLR model modifies coefficient of efficiency was found considerably significant for all models. This may be due to the successful implementation of center limit theorem.

In case bank dataset the diagnosis index of multi-collinearity was found higher in factor analysis than MLR, which signifies that factor analysis is better technique to identify multi-colinearity problem.

The F value in case of bank dataset was found significant under MLR model but F value was found very low rather in case of factor analysis was found 200 times more than the critical value, which means overall significance of all factor analysis model is higher than MLR model. The prediction plots of the two techniques (see fig. 7-fig. 9 and fig. 16- fig. 18) corroborate our discussion.

### 5.3 Interpreting of Parkinson Dataset

In case of Parkinson dataset forward model of MLR was found very low explanatory power, it is due to hetroscedasticity in stochastic error term of the model, but the full and the stepwise model was found to have 90% explanatory power of the model. In all models of factor analysis $R^2$ was found to have 60%, which is considerably sufficient for satisfactory explanatory power of the model. Moreover adjusted $R^2$ was found similar in both techniques i.e., MLR and factor analysis, due to no intrapolation.

In case of MLR models on Parkinson dataset MSE was found low and up-to the mark, which signifies that MLR technique is better technique for the extraction of structural parameters with unbiasness and low variance. On the other hand factor analysis was found having high biasness and high variance for extracting structural parameters of the model.

RMSE was found similar in all models of MLR and factor analysis which signifies the same consideration for unbiasness and variance.

The prediction power (MAE) of two models of factor analyis viz. PCR and maximum likelihood was found significant but GLS model prediction power was found considerably higher than PCR and maximum likelihood methods. On the other hand MLR prediction power was found significantly different in all three models. In case of stepwise and forward models prediction power increased more than full model.

The center limit theorem for getting efficiency of the model was found incompatible, but in case of factor analysis it was found satisfactory to the center limit theorem. Overall inn case of factor analysis modified coefficient of efficiency was found increasing.

In Parkinson dataset multi-colinearity extraction index was found higher under all models of MLR techniques except forward model. In factor analysis on the same dataset, this index was found lower than MLR model. This means MLR is better technique for diagnosing multi-colinearity particularly with full and stepwise methods.

The significance of overall model was found higher in two models of MLR viz. full and stepwise methods but in case of factor analysis, overall significance of regression model was found similar in all methods. The forward method of MLR generated considerably low F value, which means overall significance is poor than another models of both technique. The prediction plots of two techniques on Parkinson dataset is given in figure 10 to figure 12 and figure 19 to figure 21.

**6. Conclusion and Future Work**

The analysis of linear techniques (MLR and Factor Analysis) suggests that factor analysis is considerably better technique than MLR. The principal component model extracted good performance on all datasets of the study. The good performance is said on the basis of higher explanatory power, higher goodness of fit, and higher prediction power.

In diagnosis of multi-colinearity PCR model of factor analysis was found better model. However, full model of MLR also extracted satisfactory result. All other models of both the techniques were found with high explanatory power but with moderate prediction power.

All models are best fit from the point of view of linearity and unbiased ness due to moderate variance and heteroscedasticity, distribution of residual term. Their prediction power was found considerably moderate fit.

From the point of view of structural parameters and overall significance of regression model again factor analysis was found significantly up-to the mark.

From overall analysis of regression technique we can say that data with high skew ness and large structural observations should be estimated/treated with principal component model of factor analysis. The dataset with high multi-colinearity should also be treated through factors/components according to relevancy. The small dataset on the other hand should be extracted through full model of multiple regression.

The compatibility of a technique on particular dataset also depends on particular dataset's distribution of residual term of the model. In our study marketing or Parkinson dataset are having normal distribution of the residual term, on the other hand bank dataset residual term was found non normally distributed considerably. The violation of this residual assumption is affecting the prediction power for removing heteroscedastic variance of residual term. The method GLS should be adopted to estimate the structural parameters with suitable suggested forms of the regression model.

The techniques in which estimators satisfy BLUE (best, linear, unbiased, and efficient) properties of structural parameters estimates and stochastic random error term are considered better than others.

The skewness of predictors and random term in the linear regression model is creating obstacles to satisfy BLUE properties. Reducing skewness with some advance data mining tool and then comparing performance of said techniques can further enlighten us, which is an area that can be further explored.